\documentclass[aps,prapplied,reprint,superscriptaddress,floatfix]{revtex4-2}
\usepackage[utf8]{inputenc}
\usepackage[T1]{fontenc}
\usepackage{newtxtext} 
\usepackage{epsfig}
\usepackage{amsfonts}
\usepackage{amsmath}
\usepackage{amssymb}
\usepackage{amsthm}
\usepackage{color}
\usepackage{multirow}
\usepackage[normalem]{ulem}
\newcommand{\stkout}[1]{\ifmmode\text{\sout{\ensuremath{#1}}}\else\sout{#1}\fi}
\usepackage{latexsym}
\usepackage{mathrsfs}
\usepackage{natbib}
\usepackage{verbatim}
\usepackage{color}
\usepackage{booktabs}
\usepackage{dsfont}

\usepackage{mathtools}

\DeclareMathOperator{\Tr}{tr}

\newcommand{\ket}[1]{|#1\rangle}

\newcommand{\ketbra}[2]{|#1\rangle\!\langle#2|}

\newcommand{\id}{\openone}

\usepackage[colorlinks=true,linkcolor=blue,citecolor=magenta,urlcolor=blue]{hyperref}

\begin{document}


\title{Weak entanglement improves quantum communication using only product measurements}

\author{Am\'elie Piveteau}
\affiliation{Department of Physics, Stockholm University, S-10691 Stockholm, Sweden}

\author{Alastair A.\ Abbott}
\affiliation{Univ.\ Grenoble Alpes, Inria, 38000 Grenoble, France}

\author{Sadiq Muhammad}
\affiliation{Department of Physics, Stockholm University, S-10691 Stockholm, Sweden}

\author{Mohamed Bourennane}
\affiliation{Department of Physics, Stockholm University, S-10691 Stockholm, Sweden}

\author{Armin Tavakoli}
\affiliation{Physics Department, Lund University, Box 118, 22100 Lund, Sweden}


\begin{abstract}	
We show that weakly entangled states can  improve communication over a qubit channel using only separate, interference-free, measurements of individual photons. We introduce a communication task corresponding to the cryptographic primitive known as secret sharing and show that all steerable two-qubit isotropic states provide a quantum advantage in the success rate using only product measurements. Furthermore, we show that such measurements can even reveal communication advantages from noisy partially entangled states that admit no quantum steering. We then go further and consider a stochastic variant of secret sharing based on more sophisticated, yet standard, partial Bell state analysers, and show that this reveals advantages also for a range of unsteerable isotropic states. By preparing polarisation qubits in unsteerable states, we experimentally demonstrate improved success rates of both secret sharing tasks beyond the best entanglement-unassisted qubit protocol. Our results reveal the capability of simple and scalable measurements in entanglement-assisted quantum communication to overcome large amounts of noise.
\end{abstract}


\maketitle


\section{Introduction}
Entanglement is  well-known to be a crucial resource for quantum communication applications. However, not all forms of entanglement are always useful.  For instance, while entanglement that is strong enough to generate nonlocality has been found to improve noiseless classical communication beyond its conventional limitations (see, e.g.,~\cite{Cleve1997, Brukner2004, Cubitt2010, Brunner2013, Tavakoli2017, Tavakoli2020}), weaker entangled states that cannot violate any Bell inequality do not have that ability~\cite{Pauwels2022}. In contrast, if the system communicated is itself quantum, e.g.\ a qubit instead of a bit, then some weaker forms of entanglement also become useful. This can be seen in the celebrated dense coding protocol~\cite{Bennett1992} where a maximally entangled state, $\ket{\phi^+}=\frac{\ket{00}+\ket{11}}{\sqrt{2}}$, is exploited to allow a qubit message to transmit two bits instead of one bit. When the state is noisy, the isotropic state
\begin{equation}\label{eq:isotropic}
\rho_v=v \ketbra{\phi^+}{\phi^+}+\frac{1-v}{4}\id,
\end{equation}
where $v\in[0,1]$ is the visibility, achieves an advantage over an unassisted qubit message whenever $v>\frac{1}{3}$~\cite{tavakoli18, Moreno2021}. 
This coincides with the visibility at which the state becomes separable, $v_\text{sep}=\frac{1}{3}$ and it is considerably lower than the critical visibility for quantum steering, $v_\text{unsteer}=\frac{1}{2}$ \cite{renner2023compatibility, zhang2023exact}, and far below the even higher visibility needed for Bell nonlocality \cite{Designolle2023}.

However, to harness the dense coding advantage one must measure in the Bell basis $\{\Phi^+,\Phi^-,\Psi^+,\Psi^-\}$, where $\Phi^{\pm}=\ketbra{\phi^{\pm}}{\phi^{\pm}}$ and $\Psi^{\pm}=\ketbra{\psi^{\pm}}{\psi^{\pm}}$ are the projectors onto the states
$\ket{\phi^\pm}=\frac{\ket{00}\pm\ket{11}}{\sqrt{2}}$ and $\ket{\psi^\pm}=\frac{\ket{01}\pm\ket{10}}{\sqrt{2}}$. 
In optical systems, it is impossible to implement a linear optics Bell basis measurement on separate photons without the use of auxiliary photons~\cite{Lutkenhaus1999}. While dense coding experiments have been reported \cite{Mattle1996, Li2002, Barreiro2008, Williams2017, Hu2018}, implementation of the Bell basis is not expected to be scalable in optical systems in the near future. Nevertheless, it has recently been found that entanglement can yield advantages in one-shot quantum communication scenarios by using much more limited, yet much less experimentally demanding, optical measurements that are compatible with passive linear optics or even just separate single-photon measurements~\cite{Piveteau2022}. However, the schemes considered thus far come with greater requirements on the quality of entanglement, in particular needing states that violate a Bell inequality.

Here, we show that the simplest optical measurements are sufficient to reveal communication advantages from highly noisy entangled states. To this end, we introduce a version of secret sharing. In secret sharing, a secret is distributed between two parties in such a way that they must cooperate to reconstruct it~\cite{Blakley1979, Shamir1979}. This is of considerable interest for quantum cryptography. Previous protocols have achieved secret sharing in a device-dependent way, i.e.~by assuming that devices perform specific operations (see e.g.~\cite{Hillery1999, Karlsson1999, Guo2003, Tavakoli2015, Karimipour2015}). In contrast, our protocols are semi-device-independent, in the sense that all operations are uncharacterised up to the assumption that they act on qubit systems.
We prove that for our secret sharing task every steerable isotropic state enables an advantage over unassisted qubits using only product measurements on separate photons. Furthermore, we also identify unsteerable states for which an advantage is possible with such measurements. In addition,  we prove that, at the cost of using somewhat more sophisticated measurements, namely partial Bell state analysers, which require optical interference but are still feasible with standard passive linear optics \cite{Weinfurter1994, Braunstein1995}, it is possible to reveal advantages from unsteerable instances of the  isotropic state. For this, we use a stochastic modification of the original secret sharing task. Using polarisation qubits generated via spontaneous parametric down-conversion, we use both a maximally entangled state to experimentally demonstrate optimal secret sharing protocols, and unsteerable states to outperform the best possible entanglement-unassisted qubit protocols using both product measurements and partial Bell state analysers. 


\section{Deterministic secret sharing}
\begin{figure}
	\includegraphics[width=\columnwidth]{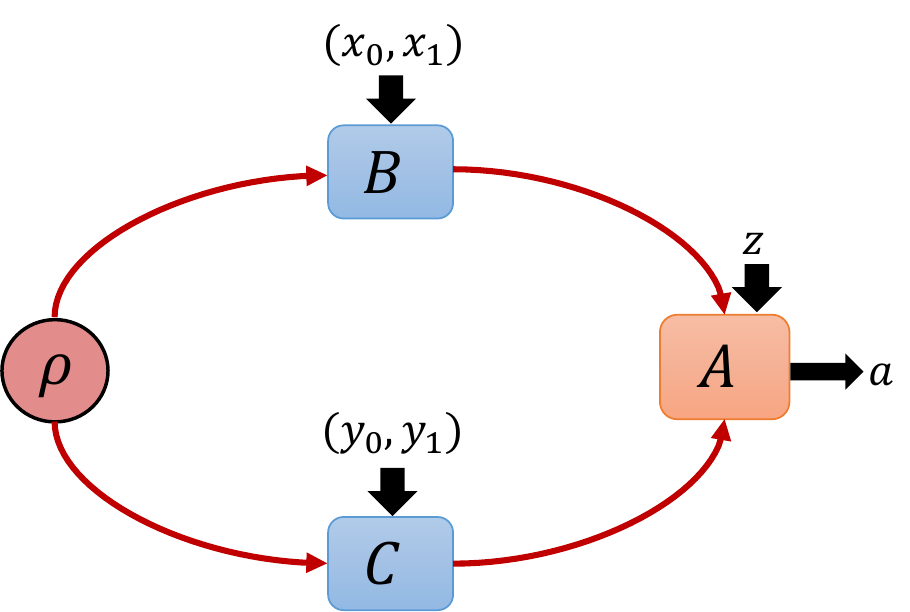}
	\caption{The secret scharing scenario. Bob and Charlie select private inputs and perform separate transformations on a shared two-qubit entangled state. They relay their respective output qubits to Alice who measures them. Her output is the secret and her input determines how it can be reconstructed collectively by Bob and Charlie.}
	\label{Schematic2}
\end{figure}

Alice wishes to generate a random secret bit $a\in\{0,1\}$ that is shared with Bob and Charlie in such a way that they individually have no knowledge of $a$ but can learn its value if they cooperate. To this end, we consider the scenario illustrated in Fig~\ref{Schematic2}. 
Bob and Charlie each privately select two uniformly random bits $x\equiv (x_0,x_1)\in\{0,1\}^2$ and $y\equiv (y_0,y_1)\in\{0,1\}^2$, respectively, and
 each communicate a message to Alice. Alice privately selects a binary input $z\in\{0,1\}$ and accordingly decodes the two incoming messages into an output $a\in\{0,1\}$. Here, $z$, which is later announced publicly, determines which bits of Bob and Charlie hold the shared secret: the success condition is $a=x_z\oplus y_z$. The average success probability of secret sharing becomes 
\begin{equation}
\begin{aligned}
& \mathcal{S}=\frac{1}{32}\sum_{x,y,z} p(a=x_z\oplus y_z|x,y,z).
\end{aligned}
\end{equation}

Naturally, the parties must have some physical limitations. In a purely classical scheme, Bob and Charlie would only share a classical random variable and they would individually send to Alice classical messages in the form of a bit respectively. Alice would use a decoding function to decide her output based on her input. However, our main focus is on quantum protocols.
 On the one hand, we are interested in the situation in which 
Bob and Charlie share no prior entanglement and simply send messages encoded in qubit states
$\beta_x$ and $\gamma_y$, respectively, while Alice can decode using a general quantum measurement $\{M_{a|z}\}_{a}$. 
The observed correlations are then given by $p_{\text{qubit}}(a|x,y,z)=\Tr\left[(\beta_x\otimes \gamma_y) M_{a|z}\right]$.
On the other hand, we also consider the situation illustrated in Fig.~\ref{Schematic2}, where Bob and Charlie may additionally share a two-qubit entangled state $\rho$ and then encode their qubit messages through local quantum channels $\Lambda_x^B$ and $\Lambda_y^C$, respectively. In this entanglement assisted case, the correlations are given by $p_\text{EAqubit}(a|x,y,z)=\Tr\left[\big(\Lambda_x^B\otimes \Lambda_y^C\left(\rho\right)\big)M_{a|z}\right]$. We now show that there exists an entanglement-assisted quantum protocol that achieves a perfect secret sharing  rate while requiring only separate measurements of the incoming quantum messages.


\section{Ideal entanglement-assisted and unassisted protocols}%
Consider a maximally entangled state $\rho=\ketbra{\phi^+}{\phi^+}$ and let Bob's and Charlie's local channels correspond to implementing the four unitaries $U^B_x=\sigma_X^{x_0}\sigma_Z^{x_1}$  and $U^C_y=\sigma_X^{y_0}\sigma_Z^{y_1}$ respectively, where $\sigma_X$ and $\sigma_Z$ are the Pauli bit-flip and phase-flip operators. This means that the two-qubit state arriving to Alice is one of the four Bell states. When $z=0$, Alice  measures $\sigma_Z\otimes \sigma_Z$, i.e., she discriminates the states $\ket{\phi^\pm}$ from the states $\ket{\psi^\pm}$. When $z=1$, she measures  $\sigma_X\otimes \sigma_X$, i.e., she discriminates the states $\ket{\phi^+} $ and $\ket{\psi^+}$ from the states $\ket{\phi^-}$ and $\ket{\psi^-}$. This gives $\mathcal{S}=1$.   

In contrast, in the scenario when entanglement is absent, it is no longer possible to succeed deterministically. We have determined the maximum value of $\mathcal{S}$ achievable with entanglement-unassisted qubits and general measurements for Alice. This is achieved using a straightforward modification of a hierarchy of semidefinite programs for bounding dimensionally-restricted quantum correlation~\cite{Navascues2015}. In order to obtain sufficiently tight upper bounds on $\mathcal{S}_\text{qubit}$, we have considered terms appearing in the first three levels of this hierarchy~\footnote{Specifically, we used what would, in standard terminology, be called the ``1+AB+AC+ABC'' level, leading to a moment matrix of size $109$.}. Up to solver precision, we obtain the upper bound $\mathcal{S}_\text{qubit}\leq \frac{3}{4}$. This bound holds even if all three parties share some pre-agreed classical randomness.

In fact, a simple classical strategy can saturate $\mathcal{S}_\text{qubit}$, thus showing both that the bound is tight and that there exists no quantum-over-classical advantage without using entanglement. 
To this end, consider that Bob and Charlie send $x_0$ and $y_0$ respectively to Alice. When $z=0$ she correctly outputs $a=x_0\oplus y_0$ and when $z=1$ she chooses $a$ at random. This leads to $S_\text{classical}=\frac{3}{4}$. We conclude that any value in the range $\frac{3}{4}<\mathcal{S}\leq 1$ implies an advantage over both the best classical and the best entanglement-unassisted quantum protocol, and hence is powered by the consumption of entanglement.


\section{Advantage from noisy entanglement}%
If, in the above ideal entanglement-assisted protocol, we substitute the maximally entangled state for the isotropic state \eqref{eq:isotropic}, we find that $\mathcal{S}=\frac{1+v}{2}$. Hence, when $v>\frac{1}{2}=v_\text{unsteer}$, product measurements reveal an entanglement-based advantage in the task. We conclude that every steerable isotropic state is a communication resource under product measurements. Furthermore, while unsteerable isotropic states do not provide an advantage, there exist other unsteerable states that are useful in the task. To this end, consider the states $\rho_v^\theta=v\ketbra{\phi^+_\theta}{\phi^+_\theta}+\frac{1-v}{4}\openone$, where $v\in[0,1]$ and $\ket{\phi^+_\theta}=\cos\theta \ket{00}+\sin\theta\ket{11}$ is a pure partially entangled state for $\theta\in(0,\pi/4]$. In \cite{Fillettaz2018}, local hidden state models have been computed for $\rho_v^\theta$. A particularly sizeable advantage is found at $\theta_{*}=0.2356$, for which the state is unsteerable when $v\lesssim 0.7325$ but achieves a secret sharing success rate $\mathcal{S}>\mathcal{S}_\text{qubit}$ whenenver $v\gtrsim 0.6878$. Hence, a significant amount of additional noise can be added to the unsteerable state before it ceases to generate a communication advantage under product measurements. Note also that applying the protocol to an arbitrary pure entangled state $\ket{\phi^+_\theta}$, we obtain $\mathcal{S}=\frac{3+\sin\left(2\theta\right)}{4}$ which exceeds the entanglement-unassisted limit if and only if the state is entangled.


\section{Stochastic secret sharing}%
We have seen that advantages in  secret sharing can be propelled by all steerable isotropic states using product measurements. Now we will see that advantages can even be obtained with unsteerable isotropic states, at the cost of switching from product measurements to standard, linear optics, partial Bell state analysers. To this end, we will consider such measurements in a stochastic version of our secret sharing task.

Consider again the scenario in Fig.~\ref{Schematic2} but now let Alice have three possible outcomes $a\in\{0,1,\perp\}$, where $a\in\{0,1\}$ corresponds to the secret bit stored in $x_{\bar{z}}\oplus y_{\bar{z}}$, where $\bar{z}=z\oplus 1$,  and $a=\perp$ is associated to a control parameter which indicates that the round is not used for secret sharing. We define that a round should be used for secret sharing exclusively when $x_z\oplus y_z=1$. Hence, the average success rate of secret sharing in the relevant rounds becomes
\begin{equation}
\begin{aligned}
& \mathcal{R}_\text{scrt}=\frac{1}{16} \sum_{z}\sum_{x_z\oplus y_z=1} p(a=x_{\bar{z}}\oplus y_{\bar{z}}|x,y,z),
\end{aligned}
\end{equation}
whereas the average probability of correctly identifying a round not used for secret sharing is $\mathcal{R}_\text{ctrl}=\frac{1}{16} \sum_{z}\sum_{x_z\oplus y_z=0} p(a=\perp|x,y,z)$. This serves as a control parameter.

We are now interested in determining the largest value of $\mathcal{R}_\text{scrt}$ for a given value of the control parameter. Note that an ideal protocol would achieve perfect secret sharing in all relevant rounds and perfectly identify all non-secret sharing rounds, i.e.~$\mathcal{R}_\text{scrt}=\mathcal{R}_\text{ctrl}=1$. This is possible with entanglement: Bob and Charlie perform the same unitaries as before and Alice does a three-outcome measurement corresponding to a projection onto  $\{\Psi^+,\Psi^-,\Phi^+ +\Phi^-\}$ for $z=0$ and onto  $\{\Phi^-,\Psi^-,\Phi^+ +\Psi^+\}$ for $z=1$.  These measurements may be seen as partial Bell state analysers, resolving two of the four Bell states, as compatible with passive linear optics \cite{Calsamiglia2001}.

In contrast, in the scenario when entanglement is absent, the trade-off between $\mathcal{R}_\text{scrt}$ and $\mathcal{R}_\text{ctrl}$ is non-trivial. Numerical exploration suggests that it is characterised by $\mathcal{R}=\frac{1}{2}\mathcal{R}_\text{scrt}+\frac{1}{2}\mathcal{R}_\text{ctrl}\leq \frac{5}{8}$ with $\mathcal{R}_\text{ctrl}\leq \frac{3}{4}$. Exploiting symmetries~\footnote{Note that $\mathcal{R}$ is invariant under i) simultaneous bit-flip of $x_0$ and $y_0$, ii) simultaneous bit-flips of $x_1$ and $y_1$ and iii) simultaneous swaps $x_0 \leftrightarrow x_1$  and $y_0\leftrightarrow y_1$. } to make the problem more tractable, we used semidefinite relaxations \cite{Rosset2019} to prove the upper bound $\frac{5}{8}$ up to solver precision. As in the deterministic case, this bound can be saturated classically: Bob and Charlie relay the bits $x_0$ and $y_0$, respectively, to Alice. If $z=1$, she outputs $a=x_0\oplus y_0$. If $z=0$, Alice outputs $a=\perp$ if $x_0\oplus y_0=0$ and $a=0$ otherwise. This gives $\mathcal{R}_\text{scrt}=\frac{3}{4}$ and $\mathcal{R}_\text{ctrl}=\frac{1}{2}$, thus saturating the entanglement-unassisted bound.

If the above entanglement-assisted protocol is performed using the isotropic state \eqref{eq:isotropic}, we find that $\mathcal{R}=\frac{3+5v}{8}$. 
Thus, we observe an advantage ascribed to entanglement whenever $\mathcal{R}>\frac{5}{8}$, which occurs whenever $v>\frac{2}{5}$. Consequently, when $v\in\left(\frac{2}{5},\frac{1}{2}\right]$, the isotropic state is both unsteerable and a communication resource with the employed decoding resources.


\section{Experiment}%
We report here on experimental implementations of the deterministic and stochastic secret sharing protocols. We implement both the ideal quantum protocols using maximally entangled states encoded in the polarisation of photons, and the protocols using weakly entangled, unsteerable, states. 
We begin by presenting the deterministic secret sharing experiment using partially entangled states $\rho_v^\theta$, before explaining how this is modified for the stochastic protocol using the simpler isotropic states $\rho_v$.

\begin{figure}
	\includegraphics[width=\columnwidth]{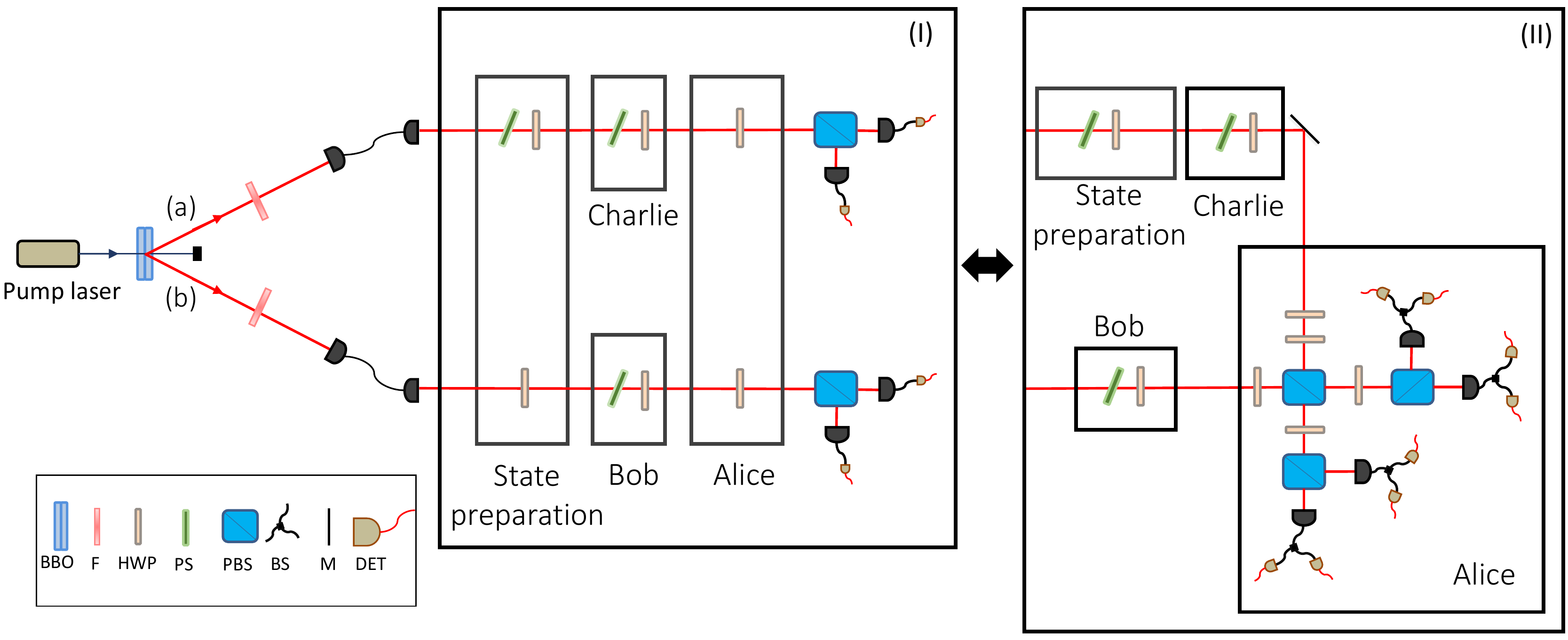}
	\caption{Experimental setup. Entangled photon pairs in spatial modes (a) and (b) are generated through the SPDC process.
For the deterministic secret sharing experiment (Box I), entangled photon pairs are generated by the same way but with a different ratio between the two modes. Isotropic states are prepared by randomly transforming the maximally entangled $\ket{\phi^{+}}$ state into one of three other Bell states using two half wave plate (HWP) and a phase shifter (PS). The unitaries of Bob and Charlie are implemented using half wave plates (HWP) and phase shifters (PS). Alice's projective measurements are implemented using HWP, polarising beam splitters (PBS) and then detected by single photon detectors (DET).
 For stochastic secret sharing experiment (Box II), isotropic states are prepared by randomly transforming the maximally entangled $\ket{\phi^+}$ state into one of the other Bell states using one half wave plate (HWP) and a phase shifters (PS). The unitaries of Bob and Charlie are implemented using half wave plates (HWP) and phase shifters (PS). Alice's partial Bell state measurements are implemented using HWP, polarising beam splitters (PBS), beam splitters (BS) and then detected by single photon detectors (DET).
}
	\label{fig:setup}
\end{figure}

To generate the entangled states, ultraviolet light centered at a wavelength of 390 nm is focused onto two 2 mm thick $\beta$ barium borate (BBO) nonlinear crystals placed in an interferometric configuration to produce photon pairs emitted into two spatial modes (a) and (b) through the second order degenerate type-I spontaneous parametric down-conversion process (SPDC).  The spectral, and temporal distinguishability between the down-converted photons is carefully removed by passing through narrow-bandwidth interference filters and quartz wedges respectively (see Fig.~\ref{fig:setup}). 
For the deterministic secret sharing protocol, we prepare the pure partially entangled state $\ket{\phi_\theta^{+}}=\cos\theta \ket{HH}+\sin\theta\ket{VV}$ by setting the polarisation of the pump laser ($\cos\theta \ket{H}+\sin\theta\ket{V}$), where $H$ and $V$ are, respectively, the horizontal and vertical photonic polarisation modes used to encode the basis states in the standard way as $\ket{0}:=\ket{H}$ and $\ket{1}:=\ket{V}$. We then prepare a depolarisation $\rho_v^\theta$ of the the state $\ket{\phi_\theta^{+}}=\cos\theta \ket{HH}+\sin\theta\ket{VV}$. To achieve the mixture with the maximally mixed state, we randomly transform, with probability $1-v$, the pure partially entangled state into one of these four orthogonal states: $\ket{\phi^+_\theta}$, $\ket{\phi_\theta^{-}}=\sin\theta \ket{HH}-\cos\theta\ket{VV}$, $\ket{\psi_\theta^{+}}=\cos\theta \ket{HV}+\sin\theta\ket{VH}$ and $\ket{\psi_\theta^{-}}=\sin\theta \ket{HV}-\cos\theta\ket{VH}$. These transformations were experimentally realised by a motorised rotation of two half wave plates and a phase shifter (see Fig.~\ref{fig:setup}). Details and experimental settings are provided in Appendix~\ref{app:B}. Alice performs product measurements in the $H$-$V$ and diagonal bases using two HWPs placed before the input of two PBS distributed on each of the two modes (a) and (b), see Fig.~\ref{fig:setup}, Box I. 

Bob's and Charlie's transformations $U^B_x$ and $U^C_y$, respectively, are performed using two HWPs and a PS to change the relative phase between the two modes (a) and (b) (see Fig.~\ref{fig:setup} and Appendix~\ref{app:B} for the settings).

We chose the visibility $v=0.72$ (with $\theta=\theta_*=0.2356$), for which the ideal state $\rho_v^{\theta_*}$ is unsteerable. Performing state tomography, we obtain a state fidelity of $0.9965 \pm 0.0007$ (see Appendix~\ref{app:D}), and we verified that the experimentally realised state is indeed unsteerable. In the implementation of the protocol the two-photon coincidences rate was one per second. The total number of events recorded is approximately $8\times 10^5$, leading to a parameter value of $\mathcal{S}=0.757 \pm 0.001$ wich is significantly above the entanglement-unassisted limit $\mathcal{S}=\frac{3}{4}$ (see Appendix~\ref{app:C}). We also performed the protocol using the pure state $\ket{\phi_{\theta_*}^+}=\cos\theta_* \ket{HH}+\sin\theta_*\ket{VV}$, with a total of $7\times 10^7$ two-photon coincidence events. We measured the state fidelity to be $0.990 \pm 0.0001$ and found a success probability of $\mathcal{S}=0.8519 \pm 0.0002$ which can be compared to the theoretical prediction of $\mathcal{S}\approx 0.8635$.

For the stochastic secret sharing protocol, we prepared the maximally entangled state $\ket{\phi^+}$ and isotropic states $\rho_v$ in a similar way (recalling that $\ket{\phi^+}=\ket{\phi^+_{\pi/4}}$). In this case the mixing with white noise, obtained by randomly transforming, with probability $1-v$, $\ket{\phi^+}$ into one of the four Bell states $\ket{\phi^\pm},\ket{\psi^\pm}$ is simplified, and was realised by a motorised rotation of one half wave plate (HWP) and a phase shifter (PS) placed on the mode (a); see Fig.~\ref{fig:setup}.
Bob's and Charlie's transformations remain the same as in the deterministic secret sharing protocol.

The polarisation measurements are performed using HWP and polarising beam splitters (PBS), beam splitters (BS) placed at the two output ports of the PBS, and then by single photon detectors (actively quenched Si-avalanche photodiodes). The partial Bell analyser (Fig.~\ref{fig:setup}, Box II) is implemented through two-photon interference, using PBS and HWPs set at $22.5^\circ$. The two-photon Hong–Ou–Mandel dip visibility is $99.2\%\pm 0.6$, where the substantial statistical error is due to the low two-photon coincidence rate used in the experiment (one per second) and a measurement time of 2400 seconds per point (see Appendix~\ref{app:G}). To switch from a Bell measurement discriminating between $\{\Psi^+,\Psi^-,\Phi^+ +\Phi^-\}$ and $\{\Phi^-,\Psi^-,\Phi^+ +\Psi^+\}$, Alice uses three HWPs placed before the first PBS (see Appendix~\ref{app:E} for the HWP settings). 
All single-detection events were registered using a VHDL-programmed multichannel coincidence logic unit, with a time coincidence window of $1.7$ ns.

To perform state tomography of isotropic states $\rho_v$ for different $v$, we performed tomography of the four Bell states generated in the randomisation procedure.
Measurements were made at a rate of one two-coincidence per second over 1400 seconds for each of the nine settings needed to perform the state tomography for each Bell state.
These results were then recombined \emph{a posteriori} at different ratios to establish the density matrices $\rho_v$.
We considered the reconstructed states $\rho_v$ for $v=0.4$ to $v=0.5$, with a step-size of $0.01$, corresponding to the resourceful but unsteerable range. 
Naturally, the reconstructed density matrices are not exactly isotropic states. 
To ensure the unsteerability of the experimentally realised states, we used the linear programming method of Ref.~\cite{Nguyen2019} which allows one to obtain a certificate of unsteerability for an arbitrary two-qubit state.
We chose to proceed during the experiment with $v=0.47$, as this represented a good balance between being below the steering threshold of $v_\text{unsteer}=\frac{1}{2}$ while allowing for good enough statistics to show a significant quantum advantage (see Appendix~\ref{app:A}). 
The fidelity of the reconstructed state with the target isotropic state for $v=0.47$ is $ 0.9983 \pm 0.0004$. 
Detailed tomography results for the density matrices, the state fidelities, and the certificates of unsteerability are presented in Appendix~\ref{app:H}.

The protocol was then carried out with the isotropic state at $v=0.47$ with the noise added by randomly changing the Bell state between each two-photon coincidence while maintaining the necessary ratio between these states. 
To obtain at most one event per change of Bell state and thus ensure the randomness and therefore unpredictability of each event, we chose to work at a rate one two-photon detection coincidence per second.
The effective measurement time per setting was $2.8$ hours. We obtained a success probability of $\mathcal{R}=0.6524 \pm 0.0004$ which significantly goes beyond the theoretical entanglement-unassisted limit of $\mathcal{R}_\text{qubit}\le\frac{5}{8}$, hence showing an advantage from unsteerable states (see Appendix~\ref{app:F}). 

The same experiment was also performed for a maximally entangled state $\ket{\phi^+}$, i.e.~without any randomisation over Bell states. This increased the rate  to 800 two-photon detection coincidences per second, with a measurement time per setting of 2 hours. 
The fidelity of the state was measured to be $0.9947 \pm 0.0009$. The observed success probability is $\mathcal{R}=0.9748 \pm 0.0001$, which is close to the ideal maximum value of $\mathcal{R}=1$, thus showing a large advantage due to entanglement.


\section{Discussion}%
\begin{figure}
	\centering
	\includegraphics[width=1\columnwidth]{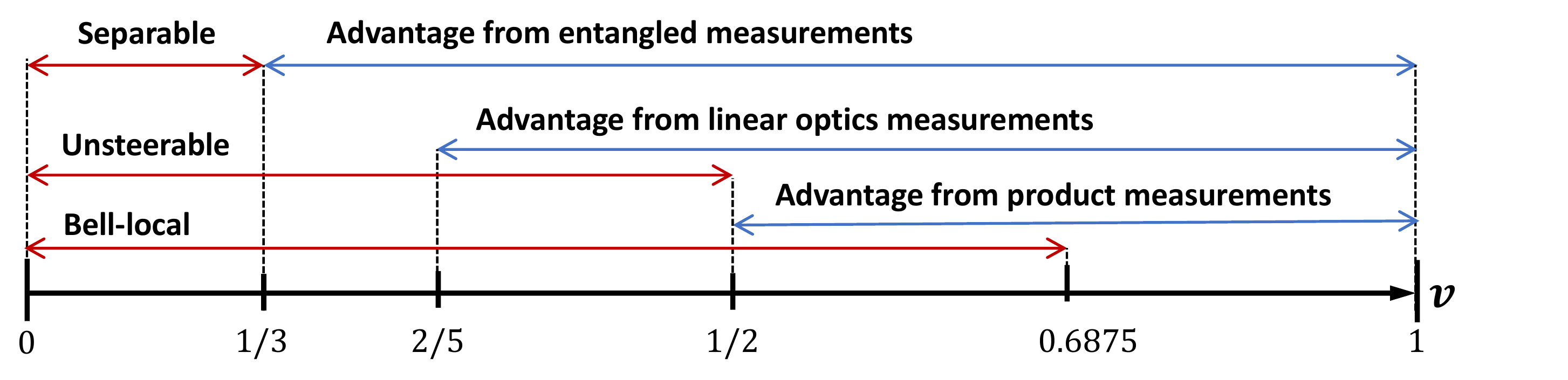}
	\caption{Nonlocal properties (red) and quantum communication advantages (blue) of the two-qubit isotropic state. The range for quantum communication advantages from general entangled measurements follows from dense coding-like protocols~\cite{tavakoli18, Moreno2021}. The ranges shown for quantum communication advantages with passive linear optics measurements and product measurements are established in this work and are upper bounds on the critical visibilities.}
	\label{FigIsotropic}
\end{figure}

In this letter we demonstrated theoretically, and confirmed experimentally, that one can obtain quantum communication advantages in a secret sharing task using weakly entangled unsteerable states while restricting only to product measurements or entangled measurements compatible with passive linear optics. Figure~\ref{FigIsotropic} summarises the ranges of noise $v$ for which quantum communication advantages are achievable with two-qubit isotropic states and the relation to its nonlocal properties. It remains an open question whether the visibilities can be further reduced both for standard linear optics measurements and product measurements. Of particular relevance is to investigate whether product measurements also can generate advantages from highly noisy multi-qubit states, as this would pave the way for experiments that go beyond small-scale demonstrations.

\begin{acknowledgments}
	A.A.A.\ and A.T.\ thank Anthony Martin and Alek Lagarrigue for discussions on an early iteration of the task presented here.
	This work  supported by the Swedish research council, the Wenner-Gren Foundation and  by the Knut and Alice Wallenberg Foundation through the Wallenberg Center for Quantum Technology (WACQT).
\end{acknowledgments}



\bibliography{bib_PrApplied}


\appendix

\section{Unsteerability of experimental states}
\label{app:A}

Recall that the isotropic state $\rho_v$ is defined as
\begin{equation}
	\rho_v = v\Phi^+ + \frac{(1-v)}{4}\id,
	\label{eq:isotropicApp}
\end{equation}
where $\Phi^{+}=\ketbra{\phi^+}{\phi^+}$.
It is well known that $\rho_v$ is unsteerable for $v\le \frac{1}{2}$ and entangled for $v> \frac{1}{3}$~\cite{Wiseman2007}.
The states prepared experimentally, however, are necessarily not exactly the isotropic states. 
To exhibit an advantage in the stochastic secret sharing task from entangled but unsteerable states, it is necessary to verify that the experimentally prepared states, as reconstructed via state tomography, are indeed unsteerable. 
The same is true for the noisy partially entangled states $\rho_v^\theta$ defined as 
\begin{equation}
	\rho_v^\theta = v \ketbra{\phi^+_\theta}{\phi^+_\theta} + \frac{(1-v)}{4}\id,
\end{equation}
where $\ket{\phi^+_\theta} = \cos\theta\ket{00} + \sin\theta\ket{11}$, which were used to show an advantage in the deterministic secret sharing protocol from unsteerable states.

While two-qubit entanglement can easily be verified by calculating its negativity, verifying unsteerability is more challenging.
Here, we use a recently developed method~\cite{Nguyen2019} that allows one to obtain certificates of unsteerability via linear programming.

In particular, the approach of Ref.~\cite{Nguyen2019} computes, for an arbitrary two-qubit state and a given direction of steering (say, from Bob to Charlie or Charlie to Bob), both upper and lower bounds, $r_\text{crit}^\text{upper}$ and $r_\text{crit}^\text{lower}$, respectively, on a ``critical radius'' $r_\text{crit}$ satisfying $r_\text{crit}^\text{lower} < r_\text{crit} < r_\text{crit}^\text{upper}$.
They show that a two-qubit state $\rho$ is steerable (in a given direction) if and only if $r_\text{crit} < 1$.
Thus, by finding $r_\text{crit}^\text{lower} \ge 1$ for both possible directions of steering, we are guaranteed a state is unsteerable.

The precision of the upper and lower bounds can be increased at the cost of computational time and memory.
On a standard desktop computer, using the CPLEX packaged provided by Ref.~\cite{Nguyen2019}, we were typically able to obtain, for the experimental states approximating $\rho_v$, a gap $r_\text{crit}^\text{upper} - r_\text{crit}^\text{lower}\approx 0.02$ with a few minutes of computation.
This can be reduced by an order of magnitude by allocating somewhat more computational time.

For the tomographically reconstructed density matrices $\tilde{\rho}_v$, for $v=0.4$ to $v=0.5$ (with a step size of $0.1$), we computed $r_\text{crit}^\text{lower}$ and found the states to be unsteerable for all values except $v=0.5$.
However, to ensure that this result is robust (recalling, of course, the experimental error in the description of $\tilde{\rho}_v$) and ensure we choose a state providing a large enough quantum advantage (which diminishes as $v$ decreases), we chose to perform the experiment with $v=0.47$.

For this state, we performed a finer analysis of the unsteerability of $\tilde{\rho}_{0.47}$, taking into account the errors in each term of the density matrix.
Adopting a conservative approach, we applied the worst case errors to all subsets of density matrix elements (up to Hermiticity) before calculating $r_\text{crit}^\text{lower}$ for the (renormalised) perturbed density matrix.
We found that, in all cases, $r_\text{crit}^\text{lower} \gtrsim 1.031$ (while for the reconstructed state $\tilde{\rho}_{0.47}$ we found $r_\text{crit}^\text{lower} \approx 1.055$), showing that the prepared state is clearly unsteerable, even when errors are taken into account.

We likewise performed a similar analysis on the tomographically reconstructed density matrix $\tilde{\rho}_v^{\theta_*}$ for $\theta_*=0.2356$ and $v=0.72$.
By applying worst case errors in the same manner we found that, in all cases, $r^\text{lower}_\text{crit}\gtrsim 1.020$, while for the reconstructed state $\tilde\rho_v^{\theta_*}$ itself we found $r^\text{lower}_\text{crit}\approx 1.031$.
As for the isotropic state, we thus see that the experimentally prepared state is clearly unsteerable, even when errors are taken into account.

\section{State preparation, transformations and measurements for the deterministic secret sharing experiment}
\label{app:B}

\subsection{State implementation}
\label{app:B1}

The setup used generates an entangled $\ket{\phi^+_\theta}=\cos\theta \ket{00}+\sin\theta\ket{11}$ state with a visibility close to 100\%. To create this state, we use an HWP to change the power distribution between the H and V crystals. When the angle of this HWP is 0, only the H-polarised crystal is pumped and when the HWP is $\frac{\pi}{2}$, only the V-polarised crystal is pumped. When the angle is $\frac{\pi}{4}$, the prepared state is $\ket{\phi^+}$. We have to select the right angle ($\theta$) value to prepare the desired state.
To create the maximally mixed state  $\frac{\id}{4}$ we randomly switch between $\ket{\phi^+_\theta}$, $\ket{\phi_\theta^{-}}=\sin\theta \ket{00}-\cos\theta\ket{11}$, $\ket{\psi_\theta^{+}}=\cos\theta \ket{01}+\sin\theta\ket{10}$ and $\ket{\psi_\theta^{-}}=\sin\theta \ket{01}-\cos\theta\ket{10}$. Indeed, these states form an orthonormal basis.

We can thereby generate $\rho_v^\theta$ for any $v$ by choosing correctly the weights with which we randomly change $\ket{\phi^+_\theta}$ to one of the other states.
The switching between these different states is done with the help of two half-wave plates and a phase plate; see Fig.~2 of the main text. 
 
These plates are motorised and switched randomly (with the desired probability) from one state to another every second. 
The number of coincidences being less than one per second, the result is to switch from one state to another randomly and independently between each pair of successive coincidences. 
The values of the angles allowing one to pass from one state to another are given in Table~\ref{tab:states_preparation}.

\begin{table}[h]
	\caption{\label{tab:states_preparation} Half wave plate rotation angles (in degrees) and phase for the preparation of the differents states.}
	\begin{ruledtabular}
		\begin{tabular}{llll} State  & $\frac{\lambda_{b}}{2}$ & $\frac{\lambda_{c}}{2}$  & Phase plate \\
		\midrule
			$\ket{\phi_\theta^{+}}$  & 0  & 0 & 0\\
				$\ket{\phi_\theta^{-}}$  & 45&45  & $\Pi$ \\
						$\ket{\psi_\theta^{+}}$  & 0  & 45 & 0\\
							$\ket{\psi_\theta^{-}}$  & 45 & 0  & $\Pi$
	\end{tabular}
\end{ruledtabular}
\end{table}


\section{Bob's and Charlie's rotations}
\label{sec:BobCharlieRotations}

As described in the main text, on inputs $x=(x_0,x_1)\in\{0,1\}^2$ and $y=(y_0,y_1)\in\{0,1\}^2$, Bob and Charlie apply the unitaries $U^B_x=\sigma_{X}^{y_{0}}\sigma_{Z}^{y_{1}}$ and $U^C_y=\sigma_{X}^{y_{0}}\sigma_{Z}^{y_{1}}$, respectively.

Table~\ref{tab:rotations} summarises explicitly the operations that Bob and Charlie must perform in inputs $x$ and $y$, as well as the Bell state expected to be received by Alice in the ideal case (assuming Bob and Charlie share a $\Phi^+$ state).
The settings to be applied to achieve each of the four different possible rotations are given in Table~\ref{tab:unitaries}.
 
\begin{table}[h]
	\caption{\label{tab:rotations} Transformation table: the operations $\sigma^{x_{0}}_{X} \sigma^{x_{1}}_{Z}$ and $\sigma^{y_{0}}_{X} \sigma^{y_{1}}_{Z}$ that need to be performed by Bob and Charlie, respectively. The expected Bell state that Alice will receive in the ideal case that Bob and Charlie share a noiseless maximally entangled state is shown in the final column.}

	\begin{ruledtabular}

		\begin{tabular}{ccccccc} $x_{0}$ & $x_{1}$ & $y_{0}$ & $y_{1}$ & Bob & Charlie & Alice's expected state\\

		\midrule

    0 & 0 & 0 & 0 & $\mathds{1}$ & $\mathds{1}$ &  $\Phi^{+}$ \\
    0 & 0 & 1 & 0 & $\mathds{1}$ & $\sigma_{X}$ &   $\Psi^{+}$ \\
    0 & 0 & 1 & 1 & $\mathds{1}$ & $-i\sigma_{Y}$ & $\Psi^{-}$ \\
    0 & 0 & 0 & 1 & $\mathds{1}$ & $\sigma_{Z}$ &  $\Psi^{-}$\\
    1 & 0 & 0 & 0 & $\sigma_{X}$ & $\mathds{1}$ &  $\Psi^{-}$ \\
    1 & 0 & 1 & 0 & $\sigma_{X}$ & $\sigma_{X}$ & $\Phi^{+}$  \\
    1 & 0 & 1 & 1 & $\sigma_{X}$ & $-i\sigma_{Y}$ & $\Phi^{-}$ \\
    1 & 0 & 0 & 1 & $\sigma_{X}$ & $\sigma_{Z}$ & $\Psi^{-}$ \\
    
   1 & 1 & 0 & 0  & $-i\sigma_{Y}$ & $\mathds{1}$ &  $\Psi^{-}$ \\
   1 & 1 & 1 & 0 & $-i\sigma_{Y}$ & $\sigma_{X}$ & $\Psi^{+}$ \\
   1 & 1 & 1 & 1 & $-i\sigma_{Y}$ & $-i\sigma_{Y}$ & $\Phi^{+}$ \\
   1 & 1 & 0 & 1 & $-i\sigma_{Y}$ & $\sigma_{Z}$ & $\Psi_{+}$ \\
    
   0 & 1 & 0 & 0 & $\sigma_{Z}$ & $\mathds{1}$ &  $\Phi^{-}$ \\
   0 & 1 & 1 & 0 & $\sigma_{Z}$ & $\sigma_{X}$ & $\Psi_{-}$ \\
   0 & 1 & 1 & 1 & $\sigma_{Z}$ & $-i\sigma_{Y}$ & $\Psi_{+}$ \\
   0 & 1 & 0 & 1 & $\sigma_{Z}$ & $\sigma_{Z}$ &  $\Phi_{+}$

		\end{tabular}
	\end{ruledtabular}
\end{table} 

\begin{table}[h]
	\caption{\label{tab:unitaries}Rotation angles (in degrees) for the unitaries for Bob's and Charlie's operations.}
	\begin{ruledtabular}
		\begin{tabular}{lll} Unitary & Phase plate & HWP  \\
		\midrule
			$\mathds{1}$ & 0 & $\Pi$ \\
            $\sigma_{X}$  & 45 & 0\\
			$-i \sigma_{Y}$  & 0 & 0 \\
			$\sigma_{Z}$  & 45 & $\Pi$
	\end{tabular}
\end{ruledtabular}
	
\end{table}


\subsection{Alice's measurements}
\label{app:B3}

Depending of her setting $z \in \{0,1\}$, Alice performs a projective measurement in the diagonal or in the $H\text{-}V$ basis. To make these measurements, she uses two HWPs $\frac{\lambda_{A1}}{2}$ and $\frac{\lambda_{A2}}{2}$ placed before the input ports of each PBS (see Fig.~2 of the main text). The values of the angles of these HWPs are given in Table~\ref{tab:z_settingsDet}.
\begin{table}[h]
	\caption{\label{tab:z_settingsDet}Rotation angles for the HWPs used to implement Alice's projective measurements, for $z=0$ and $z=1$.}
	\begin{ruledtabular}
		\begin{tabular}{lll} $z$& $\frac{\lambda_{A1}}{2}$ & $\frac{\lambda_{A2}}{2}$  \\
		\midrule
			0 & 0 & 0 \\
                1  &22.5&22.5
	\end{tabular}
\end{ruledtabular}
	
\end{table}


\section{Experimental results and error estimation for the deterministic secret sharing experiment}
\label{app:C}

Tables~\ref{tab:exp2v1} and~\ref{tab:exp2v072} list our experimental results alongside the theoretical success rates $\mathcal{S}$.

The errors were calculated using the same method as for the stochastic secret sharing experiment.

\begin{table}[h]
	\caption{\label{tab:exp2v1} Theoretical and experimental success probabilities for the partially entangled state $\ket{\phi^+_{\theta_*}}$ (with $\theta_*=0.2356$), conditioned on $z$ and then averaged to give the overall success probability $\mathcal{S}$.}
	\begin{ruledtabular}
		\begin{tabular}{llll} Parameter & Theoretical & Experimental & error \\
		\midrule
			$S^{[z=0]}$ &0.5&0.4975& 0.0005\\
	
			$S^{[z=1]}$&0.3635&0.35440&0.00009\\
			
			\midrule
			$\mathcal{S}$& 0.8635&0.8519&0.0002\\
								
	\end{tabular}
\end{ruledtabular}
	
\end{table}

\begin{table}[h]
	\caption{\label{tab:exp2v072} Theoretical and experimental success probabilities for $\rho_v^{\theta_*}$ with $v=0.72$, conditioned on $z$ and then averaged to give the overall success probability $\mathcal{S}$.}
	\begin{ruledtabular}
		\begin{tabular}{llll} Parameter & Theoretical & Experimental & error \\
		\midrule
			$S^{[z=0]}$ &0.43&0.43&0.001\\
			
			$S^{[z=1]}$&0.3317&0.327&0.002\\
			
			\midrule
			$\mathcal{S}$ & 0.7617&0.757&0.001\\
								
	\end{tabular}
\end{ruledtabular}
	
\end{table}


\section{Density matrix tomography for the deterministic secret sharing experiment}
\label{app:D}

\begin{figure}
\begin{minipage}[c]{.47\linewidth}
\includegraphics[width=0.9\columnwidth]{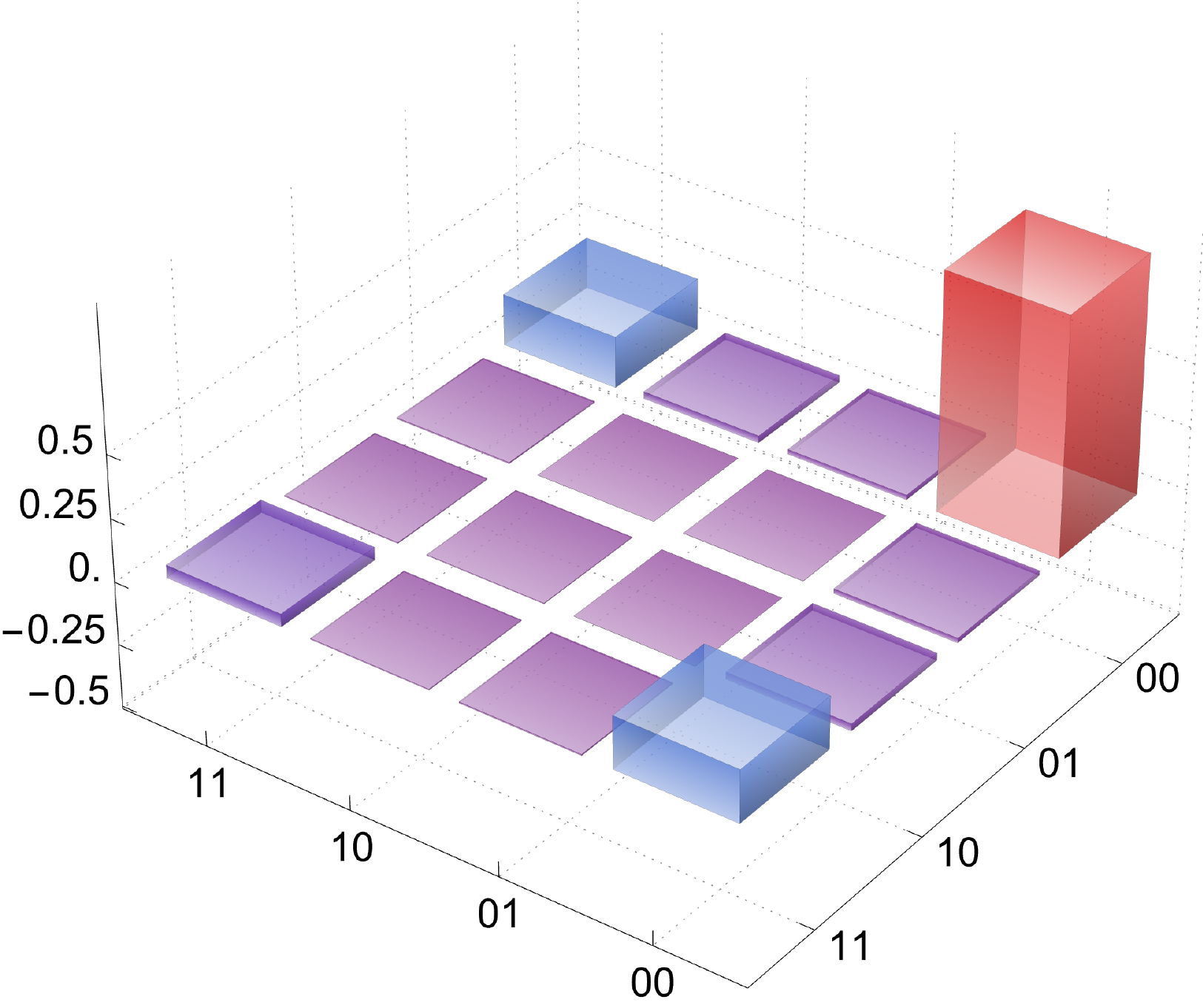}
\end{minipage} \hfill
\begin{minipage}[c]{.47\linewidth}
\includegraphics[width=0.9\columnwidth]{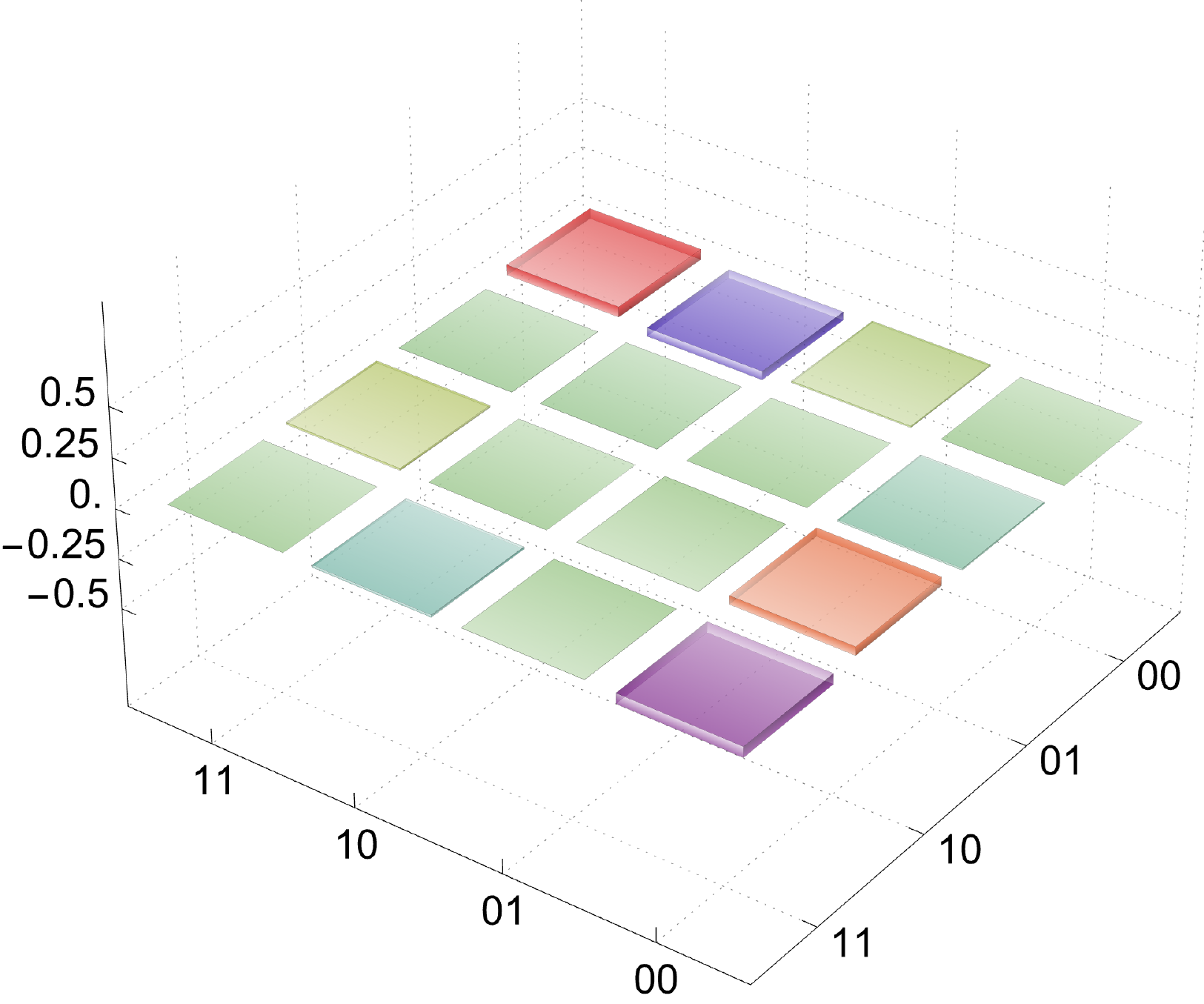}
\end{minipage}\caption{State tomography of $\rho_v^{\theta_*}$ for $v=1$; the real part of the density matrix is shown on the left, and the imaginary part on the right.}
\label{fig:imagestheta}
\end{figure}

To ensure that the prepared state is the desired one, a tomography measurement is performed for a perfect $\ket{\phi^+_{\theta_*}}$ state and for the noisy states $\rho_v^{\theta_*}$ with a visibility of $v=0.72$ (and with $\theta_*=0.2356$), which are the two states for which the experimental demonstration of the protocol was carried out. The reconstructed density matrices are shown in Figures~\ref{fig:imagestheta} and~\ref{fig:images0,72}, respectively.

\begin{figure}
\begin{minipage}[c]{.47\linewidth}
\includegraphics[width=0.9\columnwidth]{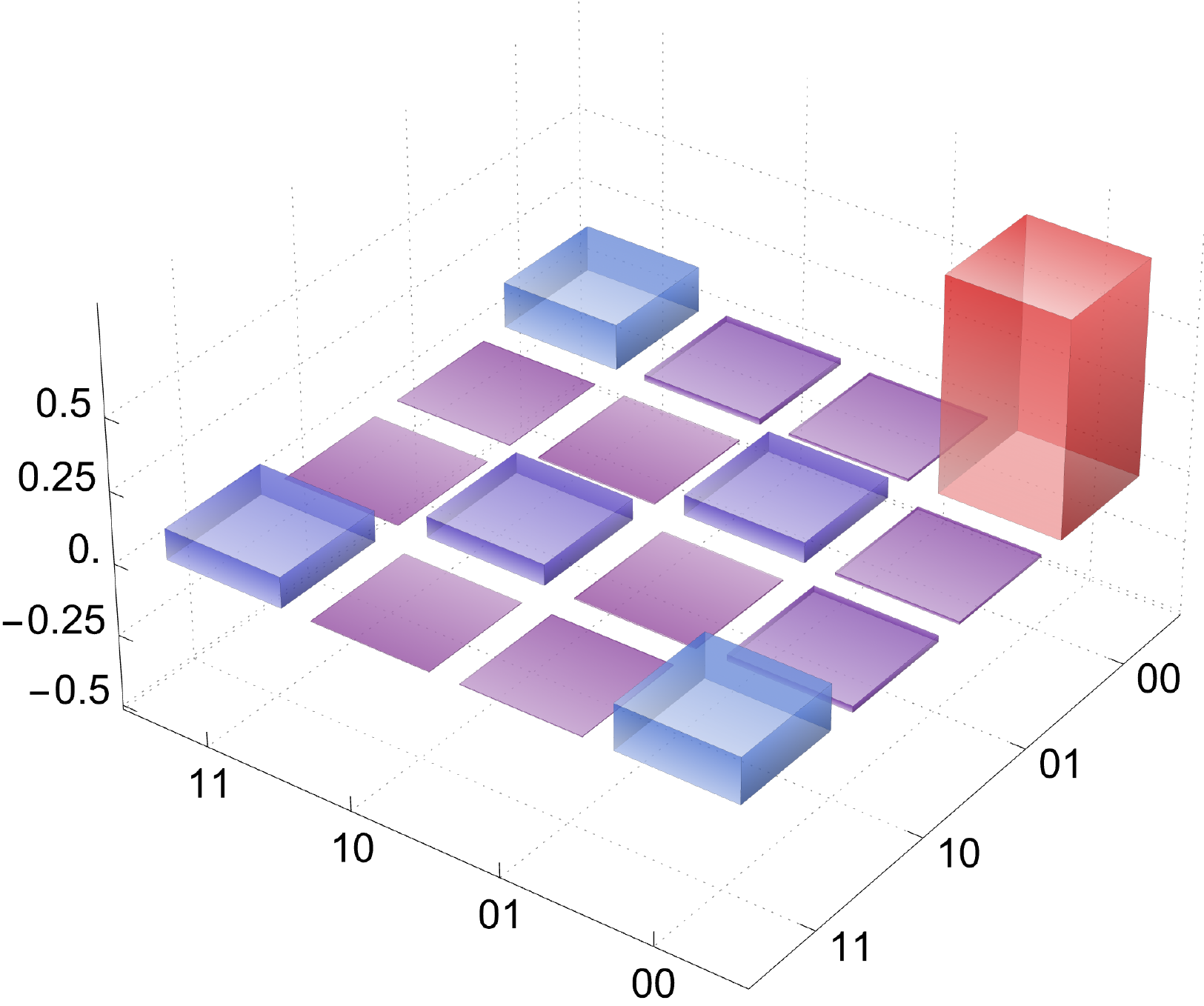}
\end{minipage} \hfill
\begin{minipage}[c]{.47\linewidth}
\includegraphics[width=0.9\columnwidth]{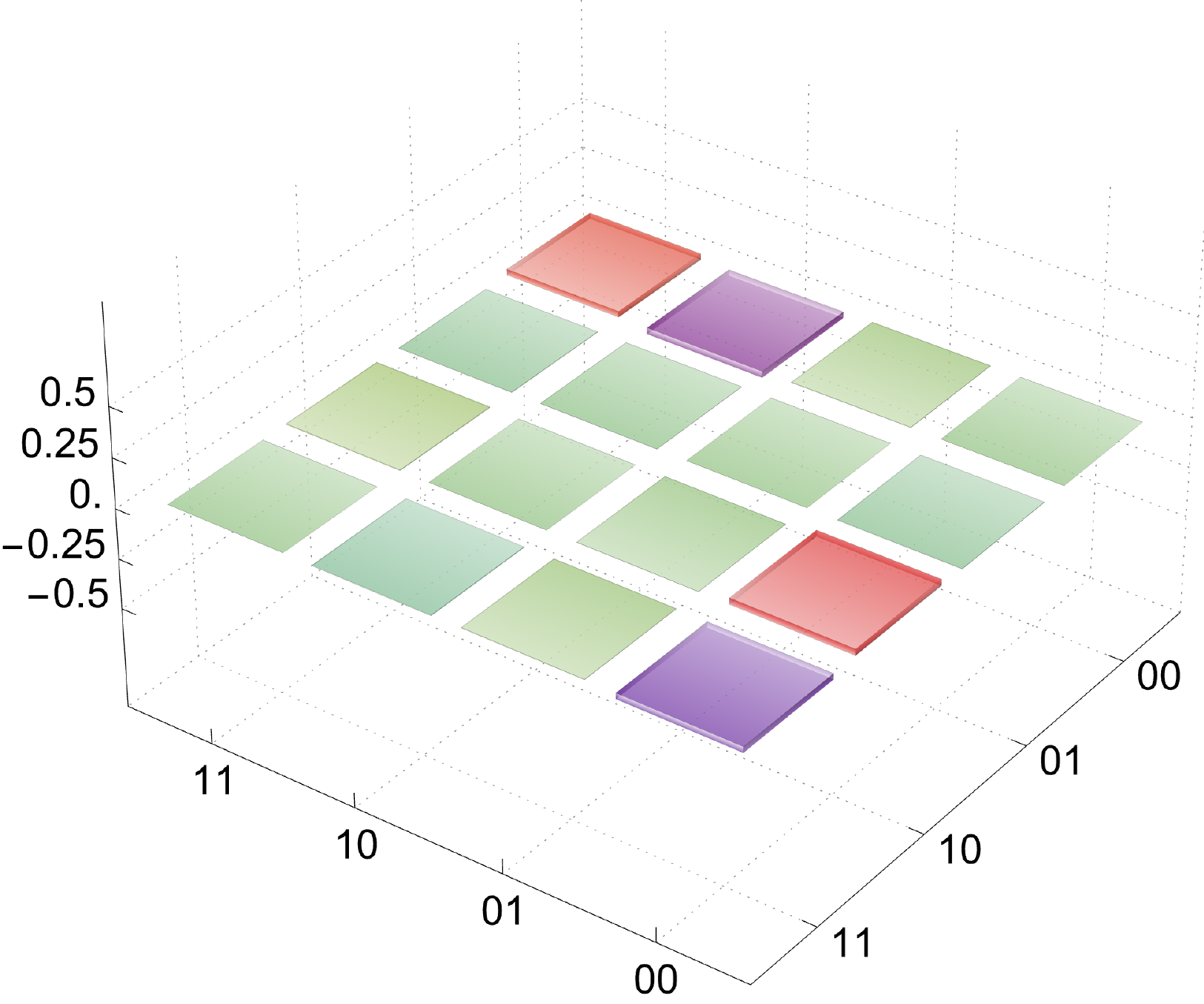}
\end{minipage}
\caption{State tomography of $\rho_v^{\theta_*}$ for $v=0.72$; the real part of the density matrix is shown on the left, and the imaginary part on the right.}
\label{fig:images0,72}
\end{figure}

The fidelity of each state obtained during the tomography with the ideal state $\rho_v^{\theta_*}$ was found to be above $0.99$. Table~\ref{tab:fidelity2} gives the fidelity for each of these states.

\begin{table}[h]
		\caption{\label{tab:fidelity2} Fidelity of measured states with the ideal state $\rho_v^{\theta_*}$ for $v=0.72$ and $v=1$.}
	\begin{ruledtabular}
		\begin{tabular}{ccc} Visibility $v$ & Fidelity with $\rho_v^{\theta_*}$ &  error \\
		\midrule
			0.72 &0.9965&0.0007\\
			
            1.00&0.990&0.001\\		
	\end{tabular}
\end{ruledtabular}
\end{table}

\section{State preparation, transformations and measurements for the stochastic secret sharing experiment}
\label{app:E}

\subsection{State implementation}
\label{app:E1}

The setup used generates an entangled $\Phi^{+}$ state with a visibility close to 100\%. 
In order to generate an isotropic state $\rho_v$ $\eqref{eq:isotropicApp}$, this state needs to be mixed with probability $(1-v)$ with the maximally mixed state $\frac{\id}{4}$.
We implement the maximally mixed state by randomly switching between the four Bell states, $\Phi^\pm,\Psi^\pm$, exploiting the fact that $\frac{\id}{4}=\frac{1}{4}(\Phi^+ + \Phi^- + \Psi^+ + \Psi^-)$.
We can thereby generate $\rho_v$ for any $v$ by choosing correctly the weights with which we randomly change $\Phi^+$ to one of the other Bell states.

The switching between these different states is done with the help of one half-wave plates and a phase plate; see Fig.~2 of the main text. 
 
These plates are motorised and switched randomly (with the desired probability) from one state to another every second. 
The number of coincidences being less than one per second, the result is to switch from one Bell state to another randomly and independently between each pair of successive coincidences. 
The values of the angles allowing one to pass from one state to another are given in Table~\ref{tab:1}.

\begin{table}[h]
	\caption{\label{tab:1} Half wave plate rotation angles (in $^\circ$) and phase for the preparation of the different Bell states.}
	\begin{ruledtabular}
		\begin{tabular}{lll} State  & $\frac{\lambda}{2}$  & Phase plate \\
		\midrule
			$\Phi^{+}$  & 0  & 0\\
				$\Phi^{-}$  & 0  & $\Pi$ \\
						$\Psi^{+}$  & 45  & 0 \\
							$\Psi^{-}$  & 45  & $\Pi$
	\end{tabular}
\end{ruledtabular}
\end{table}

\subsection{Bob's and Charlie's rotations}
\label{app:E2}

Bob's and Charlie's rotations are the same as in the stochastic secret sharing experiment; see Section~\hyperref[sec:BobCharlieRotations]{2.B} above.


\subsection{Alice's measurements}
\label{app:E3}

Depending of her setting $z \in \{0,1\}$, Alice performs one of the three-outcome partial Bell measurement $\{\Psi^{+}, \Psi^{-}, \Phi^{+}+\Phi^{-}\}$ or $\{\Phi^{-}, \Psi^{-}, \Phi^{+}+\Psi^{+}\}$ and obtains an outcome $c \in \{0, 1, \perp \}$. To make these measurements, she uses three HWPs $\frac{\lambda_{A1}}{2}$, $\frac{\lambda_{A2}}{2}$ and $\frac{\lambda_{A3}}{2}$, with the first two ($A1$ and $A2$) placed before one of the input ports of the PBS, and last one ($A3$) placed before the other input port (see Fig.~2 of the main text). The values of the angles of these HWPs are given in Table~\ref{tab:z_settingsStoch}.
\begin{table}[h]
	\caption{\label{tab:z_settingsStoch}Rotation angles for the HWPs used to implement Alice's partial Bell state measurements, for $z=0$ and $z=1$.}
	\begin{ruledtabular}
		\begin{tabular}{llll} $z$& $\frac{\lambda_{A1}}{2}$ & $\frac{\lambda_{A2}}{2}$ & $\frac{\lambda_{A1}}{2}$ \\
		\midrule
			0 & 0 & 45 & 0 \\
                1 & 0 &22.5&22.5
	\end{tabular}
\end{ruledtabular}
	
\end{table}

\section{Experimental results and error estimation for the isotropic states $\rho_v$}
\label{app:F}

Tables~\ref{tab:v1} and~\ref{tab:v047} list our experimental results alongside the theoretical success rates $\mathcal{R}$ with the related errors.

\begin{table}[h]
	\caption{\label{tab:v1} Theoretical and experimental success rate for the maximally entangled state $\Phi^+$, conditioned on $z$ and then averaged to give the overall success rate $\mathcal{R}$.}
	\begin{ruledtabular}
		\begin{tabular}{llll} Parameter & Theoretical & Experimental & error \\
		\midrule
			$R^{[z=0]}_\text{ctrl}$ &1&0.9900& 0.0002\\
			$R^{[z=0]}_\text{scrt}$& 1& 0.9811&0.0003\\
			$R^{[z=1]}_\text{ctrl}$&1&0.9795&0.0002\\
			$R^{[z=1]}_\text{scrt}$&1&0.9485&0.0003\\
			\midrule
			$\mathcal{R}$& 1&0.9748&0.0001\\
								
	\end{tabular}
\end{ruledtabular}
	
\end{table}

\begin{table}[h]
	\caption{\label{tab:v047} Theoretical and experimental success rates for the isotropic state $\rho_v$ with $v=0.47$, conditioned on $z$ and then averaged to give the overall success rate $\mathcal{R}$.}
	\begin{ruledtabular}
		\begin{tabular}{llll} Parameter & Theoretical & Experimental & error \\
		\midrule
			$R^{[z=0]}_\text{ctrl}$ &0.735&0.7238&0.0007\\
			$R^{[z=0]}_\text{scrt}$&0.6025&0.6086&0.0009\\
			$R^{[z=1]}_\text{ctrl}$&0.735&0.6613&0.0007\\
			$R^{[z=1]}_\text{scrt}$&0.603&0.6161&0.001\\
			\midrule
			$\mathcal{R}$& 0.66875&0.6524&0.0004\\
								
	\end{tabular}
\end{ruledtabular}
	
\end{table}

To calculate these errors, following \cite{PhysRevLett.125.080403} we consider errors originating from the measurement side only. 
To reduce experimental errors in the measurements, we used computer controlled high precision motorised rotation stages, for Bob, Charlie and the generation of Werner states, to set the orientation of the wave-plates with a repeatably high precision of $0.02 ^{\circ}$. 
The use of different settings $(x,y)$ induces a systematic error, which we estimate using Monte Carlo simulation. 
We assume that the wave plates' setting errors are normally distributed with a standard deviation of $0.02^{\circ}$. 
Alice's wave plates are not motorised. 
In order to reduce systematic errors in the angle settings, all settings for $z=0$ are measured, then all settings for $z=1$, so the results are initially presented separately for each case. We estimate the accuracy of the angles of these WPs to be half a degree.
This, together with the Poissonian error in photon counting statistics, comprises the final error reported here. 
Due to inefficiency in the single photon detectors, the photons are detected  randomly  and  their counting  is Poissonian. 
To decrease Poissonian counting error, we  have chosen for $v=1$ a measurement time  of  two hours  for every setting and collected approximately 93 million events.
To guarantee that both parties receive single qubits, we worked at a low rate ($\approx 800$ coincidence per sec) to suppress higher order coincidences to almost $0.9$ per sec.  
For $v=0.47$, we have collected around one million two-photon coincidences and the low rate renders the higher order coincidences completely negligible.


\section{Two-fold Hong-Ou-Mandel dip visibility}
\label{app:G}

\begin{figure}[h]
	\centering
	\includegraphics[scale=0.5]{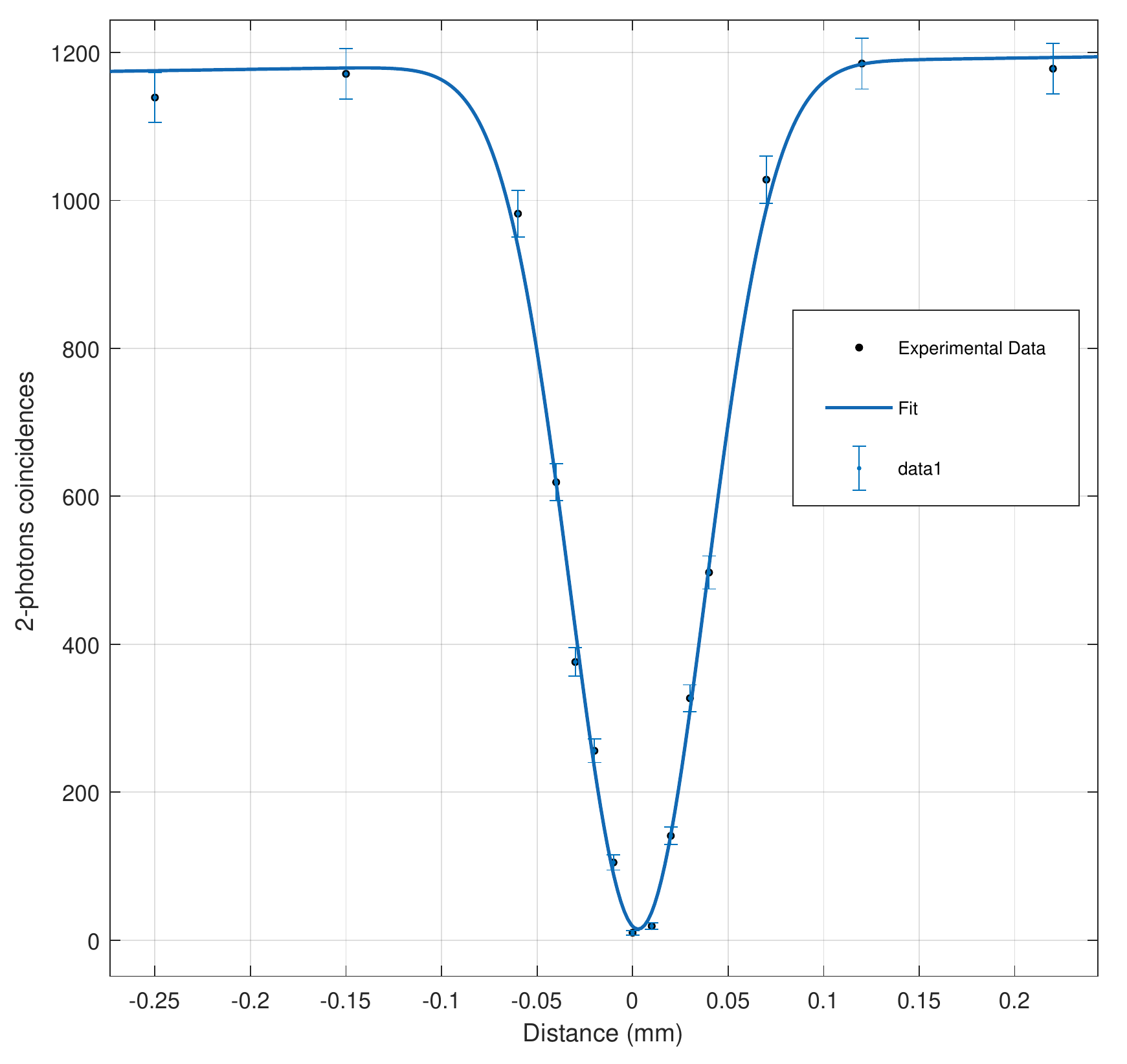}
	\caption{Two-fold Hong-Ou-Mandel dip. The plot displays the two-fold photon counting coincidences versus the delay (the path  difference between the two arms). The error bars indicate the Poissonian photon counting error statistics. The data is fitted with a Gaussian curve.}\label{fig:HOM}
\end{figure}

Bell state measurements are implemented through  two-photon interference, using PBS and HWP plates set at 22.5°. The photons are detected by Si avalanche photodiodes and the coincidences are registered with an eight channel multifold coincidence counting unit. 
This Bell analyser consists of coherent interference at a polarising beam splitter (PBS). To achieve the necessary indistinguishability of the photons, due to their arrival times, we adjusted the path length of one of the photons using a delay line~\cite{HOM}. In Figure~\ref{fig:HOM}, the coincidences between the detectors versus the delay path length is shown. The zero delay corresponds to a maximal overlap (maximum indistinguishability). The interfering  photons bunch (i.e., they both exit in the same output arm of the PBS) causing the coincidences to vanish. The measured visibility of the two-fold Hong-Ou-Mandel dip is $99.2\% \pm 0.6$, with a coincidence rate outside the dip around $0.9224$ coincidences per second, identical to the rate at which we performed the experiment for $v=0.47$.

\section{Density matrix tomography for stochastic secret sharing}
\label{app:H}

To ensure that the prepared state is the desired one, a tomography measurement is performed for a perfect $\Phi^{+}$ state and for isotropic states $\rho_v$ with a visibility between $v=0.4$ and $v=0.5$. The reconstructed density matrices obtained for the cases of $v=1$ (i.e., the ideal case of $\rho_1 = \Phi^+$) and $v=0.47$, which are the two states for which the experimental demonstration of the protocol was carried out, are presented in Figures~\ref{fig:images1,00} and~\ref{fig:images0,47}, respectively.

\begin{figure}
\begin{minipage}[c]{.47\linewidth}
\includegraphics[width=0.9\columnwidth]{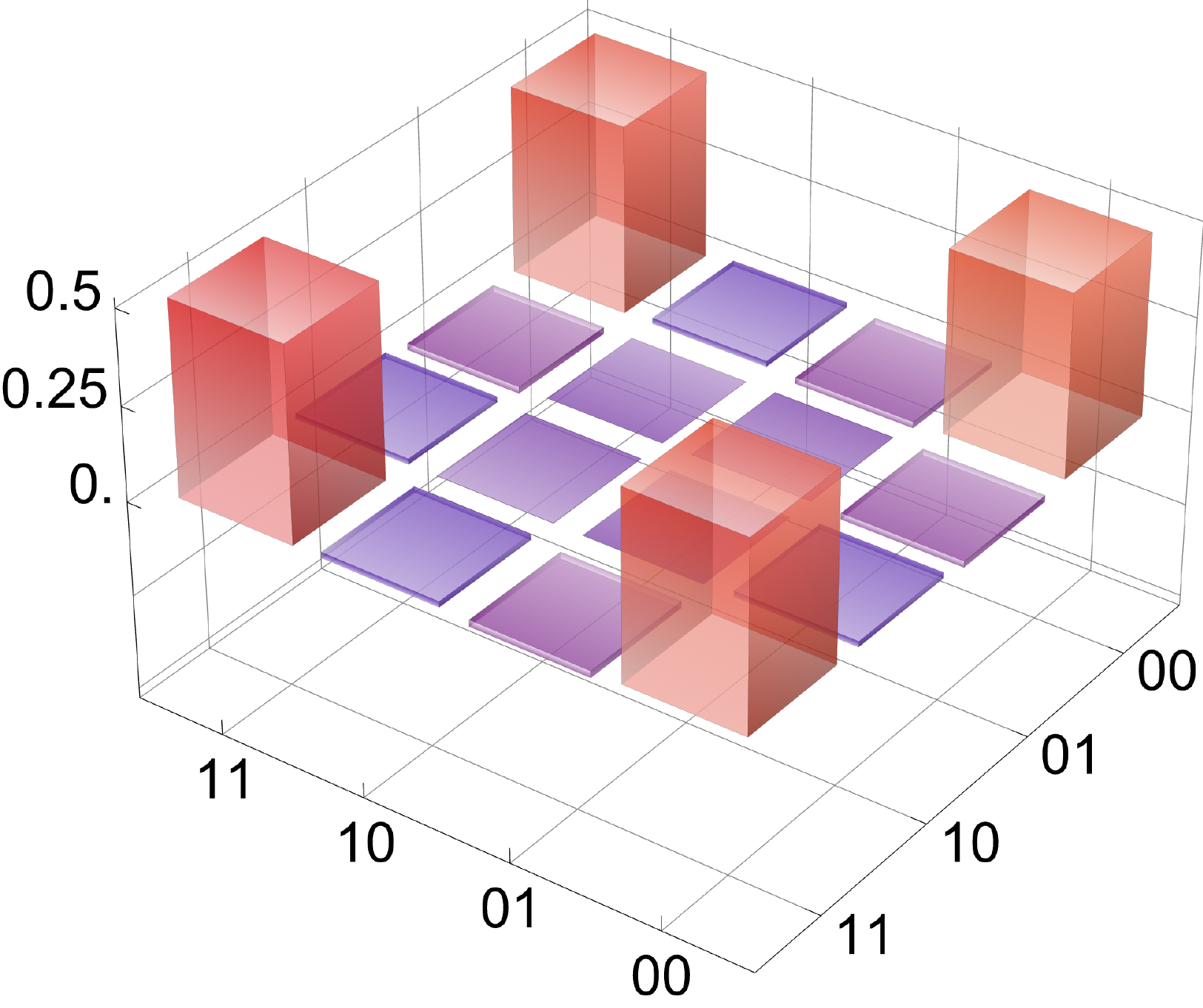}
\end{minipage} \hfill
\begin{minipage}[c]{.47\linewidth}
\includegraphics[width=0.9\columnwidth]{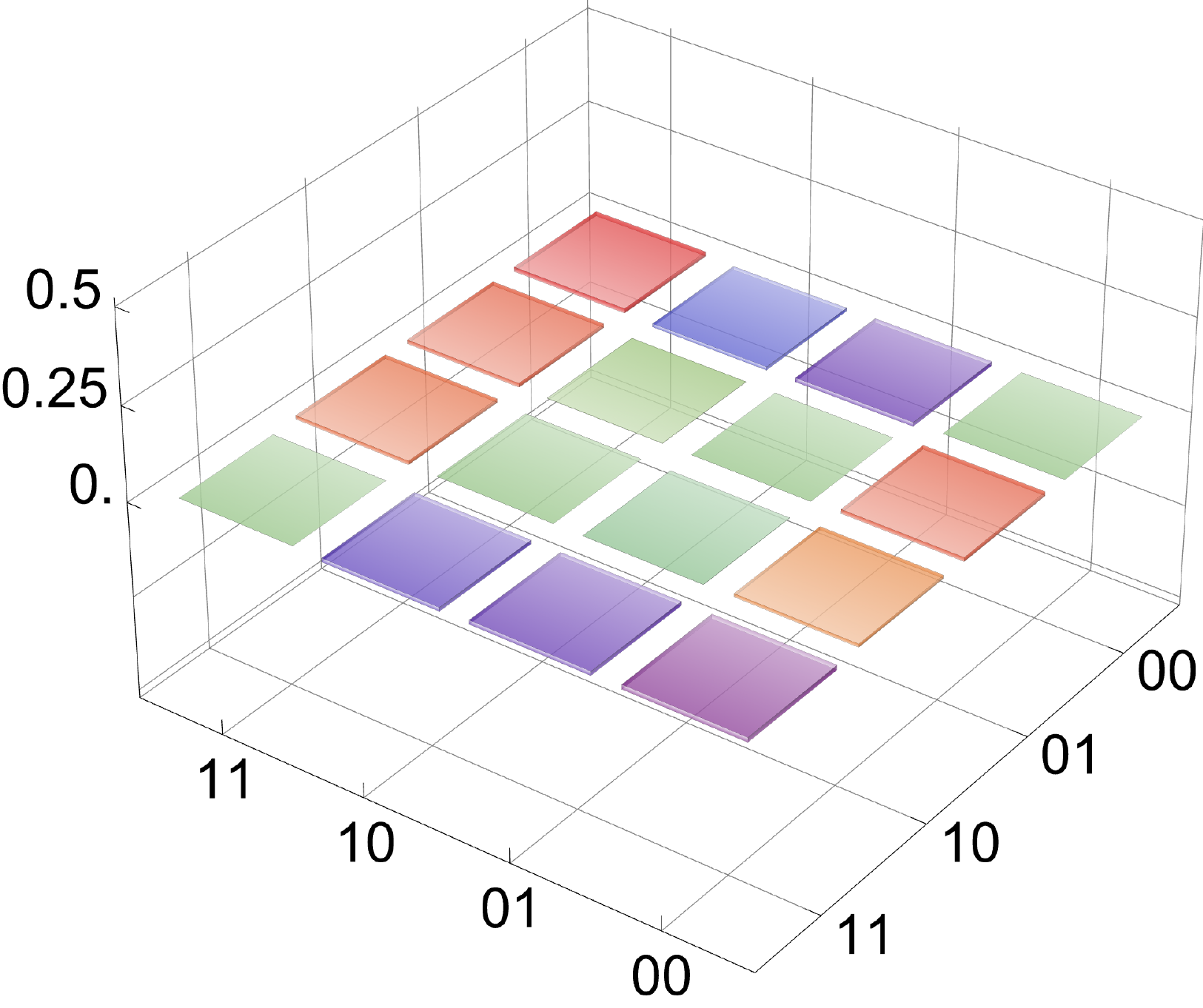}
\end{minipage}
\caption{State tomography of $\rho_v$ for $v=1$; the real part of the density matrix is shown on the left, and the imaginary part on the right.}
\label{fig:images1,00}
\end{figure}

\begin{figure}
\begin{minipage}[c]{.47\linewidth}
\includegraphics[width=0.9\columnwidth]{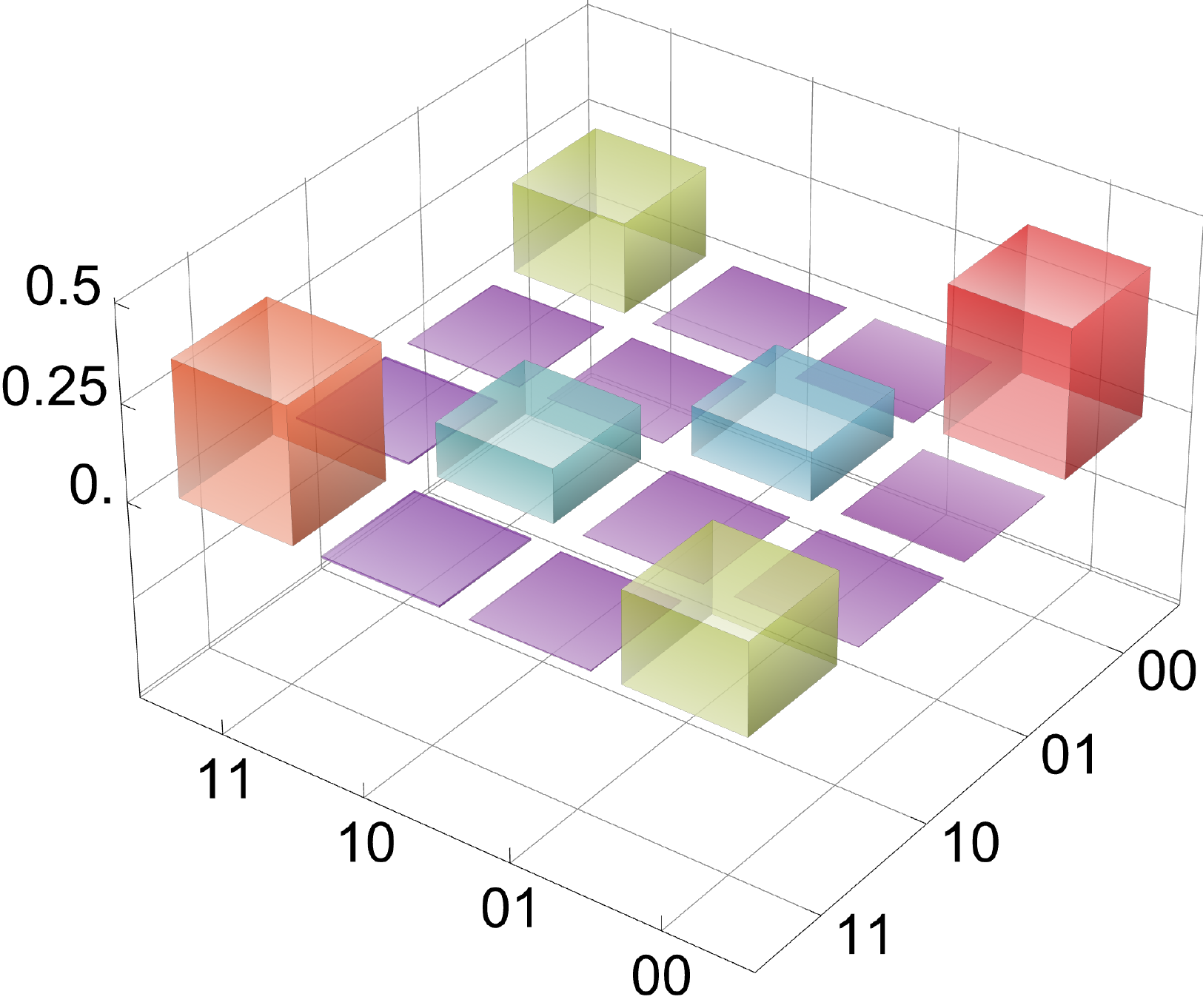}
\end{minipage} \hfill
\begin{minipage}[c]{.47\linewidth}
\includegraphics[width=0.9\columnwidth]{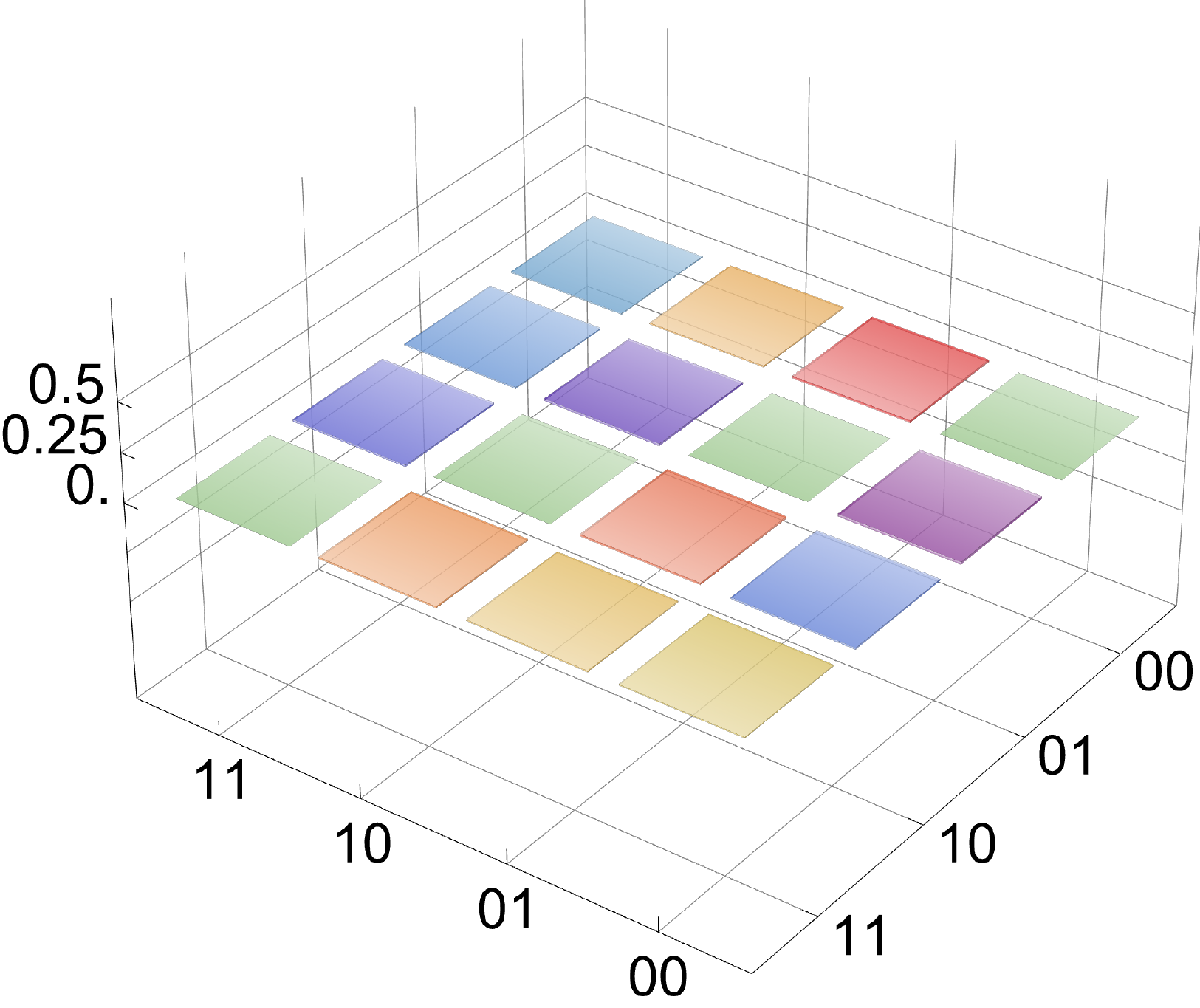}
\end{minipage}
\caption{State tomography of $\rho_v$ for $v=0.47$; the real part of the density matrix is shown on the left, and the imaginary part on the right.}
\label{fig:images0,47}
\end{figure}

The fidelity of each state obtained during the tomography with the ideal isotropic state $\rho_v$ was found to be between $0.9947$ and $0.9983$. Table~\ref{tab:fidelity} gives the fidelity for each of these states.

\begin{table}[h]
		\caption{\label{tab:fidelity} Fidelity of each measured state with the ideal isotropic state $\rho_v$.}
	\begin{ruledtabular}
		\begin{tabular}{ccc} Visibility $v$ & Fidelity with $\rho_v$ &  error \\
		\midrule
			0.4 &0.9982&0.0004\\
			0.41&0.9982 &0.0004\\
			0.42 &0.9983 &0.0004\\
            0.43&0.9983&0.0004\\
            0.44&0.9983&0.0004\\
            0.45&0.9983&0.0004\\
            0.46&0.9983 &0.0004\\
            0.47&0.9983&0.0004\\
            0.48&0.9983&0.0004\\
            0.49&0.9983&0.0004\\
            0.50&0.9983&0.0004\\
            1.00&0.9947&0.0009\\		
	\end{tabular}
\end{ruledtabular}
\end{table}

\end{document}